\documentclass[aps,prb,superscriptaddress,floatfix,longbibliography, 12pt]{revtex4-2}

\usepackage{amsmath,amssymb} 
\usepackage{bm} 
\usepackage{graphicx} 
\usepackage{comment} 
\usepackage{amsmath}
\usepackage{amssymb}
\usepackage{mathtools}
\usepackage{bbm}
\usepackage{longtable} 

\usepackage{booktabs}
\usepackage{textcomp} 
\usepackage{float}

\usepackage{parskip} 
\usepackage{listings} 

\usepackage{fouriernc}
\usepackage{tgschola}
\usepackage[T1]{fontenc}

\usepackage{enumitem}
\setlist{noitemsep,leftmargin=*,topsep=0pt,parsep=0pt}

\usepackage{xcolor} 
\definecolor{lightgray}{gray}{0.6}
\definecolor{medgray}{gray}{0.4}

\usepackage{hyperref}
\hypersetup{
colorlinks=true,
urlcolor= blue,
citecolor=blue,
linkcolor= blue,
}

\newif\ifptitle
\newif\ifpnumber
\newcounter{para}
\newcommand\ptitle[1]{\par\refstepcounter{para}
{\ifpnumber{\noindent\textcolor{lightgray}
{\textbf{\thepara}}\indent}\fi}
{\ifptitle{\textbf{[{#1}]}}\fi}}

\newcommand{\mytitle}{Real-Time Computational Visual Aberration Correcting Display Through High-Contrast Inverse Blurring}
\begin{document}

\title{\normalfont\itshape\mytitle.}

\author{Akhilesh Balaji}
\email[]{akhilesh.balaji.bangalore@gmail.com}
\affiliation{Neev Academy, Bengaluru, Karnataka, 560037 India}
\affiliation{Ashoka University, Rajiv Gandhi Education City, National Capital Region P.O. Rai, Sonepat, Haryana, 131029 India}

\author{Dhruv Ramu}
\email[]{dhruvramu@gmail.com}
\affiliation{Neev Academy, Bengaluru, Karnataka, 560037 India}
\affiliation{Ashoka University, Rajiv Gandhi Education City, National Capital Region P.O. Rai, Sonepat, Haryana, 131029 India}

\date{\today}

\begin{abstract}
\noindent This paper presents a framework for developing a live vision-correcting display (VCD) to address refractive visual aberrations without the need for traditional vision correction devices like glasses or contact lenses, particularly in scenarios where wearing them may be inconvenient. We achieve this correction through deconvolution of the displayed image using a point spread function (PSF) associated with the viewer’s eye. We address ringing artefacts using a masking technique applied to the prefiltered image. We also enhance the display's contrast and reduce color distortion by operating in the YUV/YCbCr color space, where deconvolution is performed solely on the luma (brightness) channel. Finally, we introduce a technique to calculate a real-time PSF that adapts based on the viewer's spherical coordinates relative to the screen. This ensures that the PSF remains accurate and undistorted even when the viewer observes the display from an angle relative to the screen normal, thereby providing consistent visual correction regardless of the viewing angle. The results of our display demonstrate significant improvements in visual clarity, achieving a structural similarity index (SSIM) of 83.04\%, highlighting the effectiveness of our approach.
\end{abstract}
\maketitle
\vspace{2cm}
\newpage
\section{\label{sec:Intro}Introduction}

\ptitle{The Prevalence of Optical Defocus} Today, 2.2 billion people have near or distance visual aberrations across all ages~\cite{Visionimpairmentblindness_2023}. Visual impairment is not just a health problem; annual global costs associated with visual impairment based treatment exceed 411 billion USD~\cite{steinmetz2021causes}.
Myopia is visual impairment wherein objects at a relatively far distance are not visible (nearsightedness) due to the convergence of light rays behind the retina. It is expected to affect 50\% of the world population by 2050. In hyperopia, light rays converge before the retina, resulting in farsightedness~\cite{holden2016global}.

\ptitle{The Drawbacks of Glasses/Contact Lenses} Traditional solutions to defocus are spectacles and contact lenses; they do not provide perfect solutions to the widespread visual aberration problem. Spectacles are incompatible with virtual and augmented reality platforms as they cannot be worn in conjunction with a VR headset or 3D anaglyph glasses. Blurry or dusty eyeglasses may cause further eye strain, aggravating the visual aberration they were diagnosed with. The use of contact lenses to solve this may lead to the progressive dryness of eyes~\cite{markoulli2017contact}. When afflicted by hyperopia, driving a car with glasses is necessary only for the clarity of the Head-Up Display---if this can, for instance, incorporate VCD technology, glasses for hyperopia are no longer required while driving.

\begin{figure}[H]
    \centering
    \includegraphics[width=\linewidth]{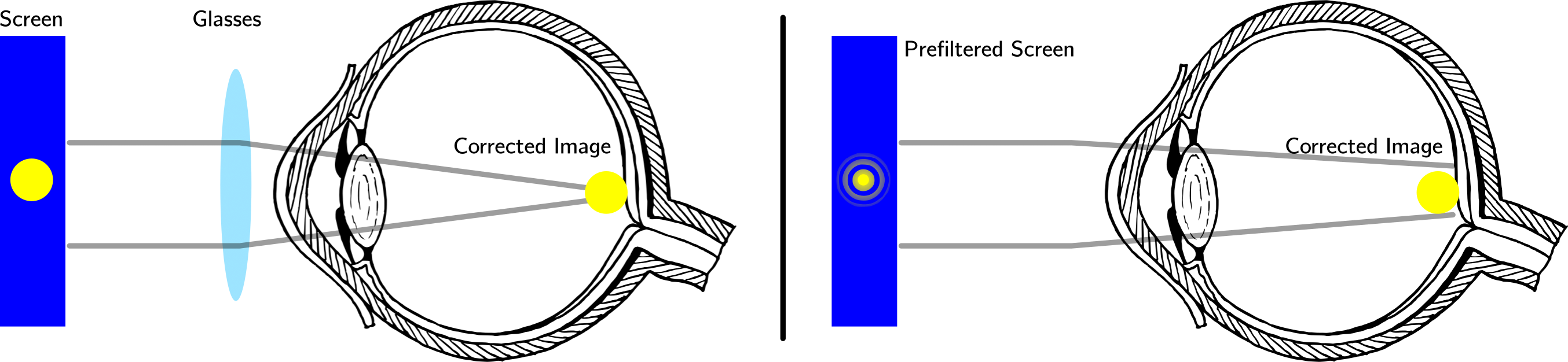}
    \caption{The eye percieves an on-screen image (blue). Left (a): Rays converge on the retina after passing through the glass lens, resulting in a clear image; Right (b): In a VCD, there are no glasses, but the on-screen image is pre-distorted---despite the rays not converging on the retina, the image appears clear.}
    \label{fig:representation}
\end{figure}

\subsection*{Problem Statement}
Glasses and contact lenses provide solutions to lens-based visual aberrations, but for higher-order aberrations, lens manufacturing becomes difficult or heavy / asymmetrical lenses may have to be dealt with. Thus, we `shift' the glasses computationally onto the screen, as seen in Figure~\ref{fig:representation}, eliminating the need for lenses to correct the aberration.

\section{Background}
The problem of vision correcting displays has primarily been addressed, in recent times, by the creation of light field displays. A light field captures the exact configuration of electromagnetic waves, complete with phase and amplitude descriptions, in a given optical system. In Huang's 2014 paper~\cite{Huang:2014}, we see that the light field surrounding a display as perceived by a visually aberrated individual can be computed by passing the light field through computational lenses to simulate the aberration and its correction. The physical generation of this light field is achieved through a \emph{light field display}---Huang et al. use pinholes placed over a pre-distorted image, such that the diffracted individual light fields emerging from each pinhole interfere with each other to produce the correct light field. This naturally implies the existence of a hardware component to achieve the light field propagation, which would have significant cost associated with it if it were to be deployed on a large-scale basis. Likewise, the authors state that the pre-processing procedure is computationally expensive and requires high processing time.

Alternatively, one may use a holographic approach~\cite{kim2021vision}. In a traditional hologram, the light field is captured on a photosensitive film by means of a light source passing over the objects to be recorded that interferes with a second coherent light beam used to capture phase difference. When exposed to the original coherent light beam once more, the light field is regenerated around the film and the three-dimensional scene is captured~\cite{xue2014multiplexing}. However, upon placing an appropriate lens simulating myopia or hyperopia over the scene, the blurred (or de-blurred, in the case of a visually aberrated viewer) nature of the light field is also captured. This allows for vision correction. However, this only works for still images, and techniques for temporally evolving holograms are still under research~\cite{blanche2010holographic}.

Thus, it can be seen that various approaches---such as light field and holographic displays---have been applied to tackle this problem. We address their limitations in our work, where we adopt the technique in \cite{alonso2003image}, wherein no hardware component is required to deblur an image. The techniques we use for optimization and colour restoration may be extended to any kind of deconvolution algorithm. While the method proposed by Alonso and Barreto is also computationally expensive, we propose methods of optimizing these algorithms to create a \emph{real-time} vision correcting display, which the user can interact with.

\section{Methodology}
\subsection{General Overview}
\begin{figure}
    \centering
    \includegraphics[width=4.8in]{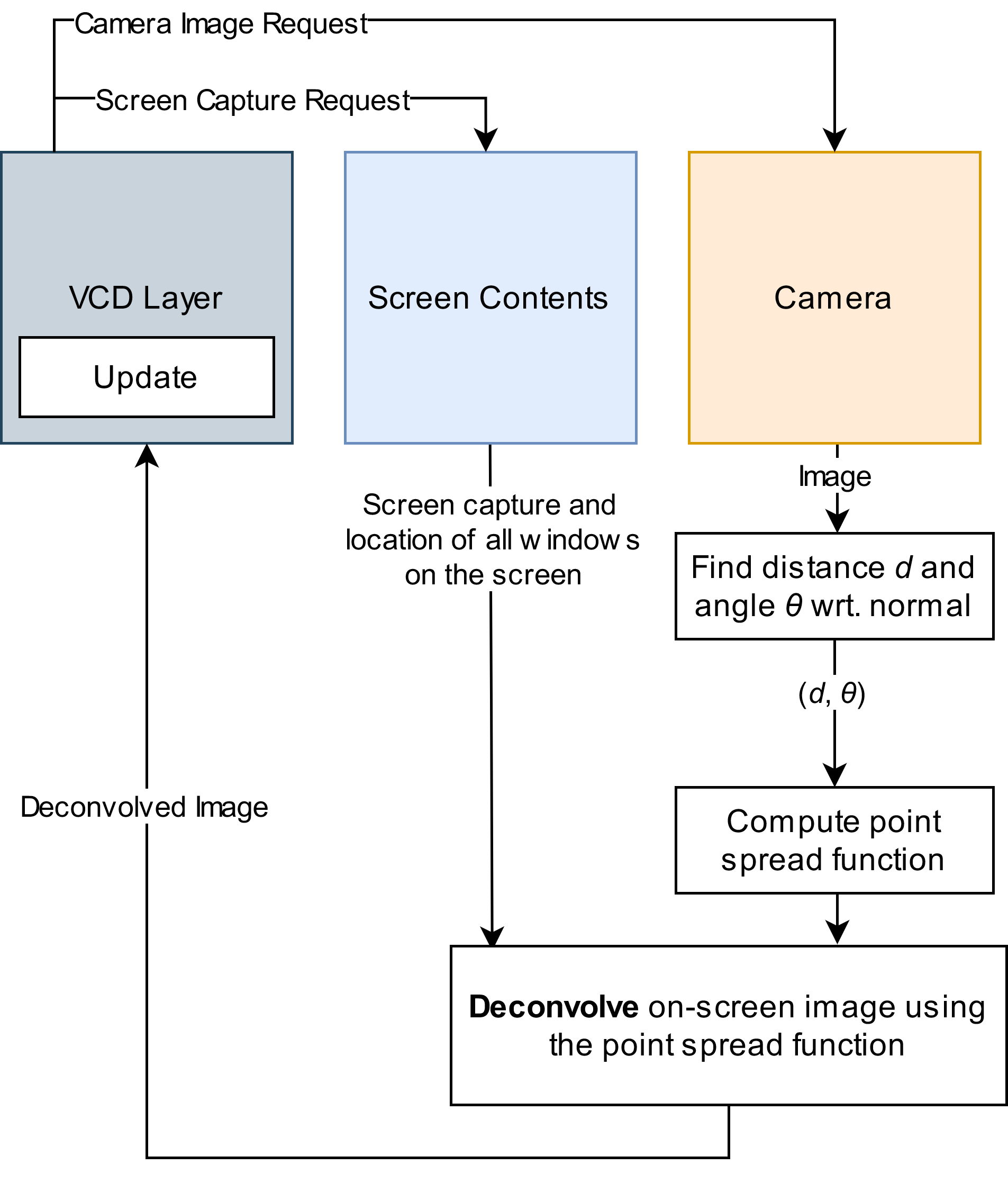}
    \caption{Flowchart detailing the basic design.}
    \label{fig:flowchart}
\end{figure}
As can be seen in Figure~\ref{fig:flowchart}, we take snapshots of the on-screen image at small intervals of time, deconvolve these according to the position of the viewer estimated at that point in time, and display this on the screen, replacing the original image fed by the screen contents.

\subsection{Theory of Vision Correction}

We use a fundamental operation in image processing, that is convolution, to simulate the blurring of the eye. The eye is characterized by a blur kernel that is represented in the point spread function (PSF; or optical transfer function, OTF) associated with it. We define the PSF to be:
\begin{align}
k(x,y)\coloneqq\begin{cases}
    \chi & \text{if \(\sqrt{x^2 + y^2} \leq r\)}, \\
    0 & \text{otherwise}.
\end{cases}
\end{align}
Here, \(\chi\) is chosen such that:
\[
\iint_{-\infty}^{+\infty} k(x,y)\;\;\mathrm{d}x\;\mathrm{d}y = V_k = 1;
\]
that is, the PSF integrates to unity such that no light is lost, nor is it gained. The image brightness is conserved. This is called normalization. This integral becomes a discrete summation when dealing with images---discrete functions may thus be represented as matrices or 2D arrays. The PSF model, as obviated above, is that of a disk, which best models defocus in the human eye, due to spurious resolution induction, as will shortly be seen.

\begin{figure}
    \centering
    \includegraphics[width=0.75\textwidth]{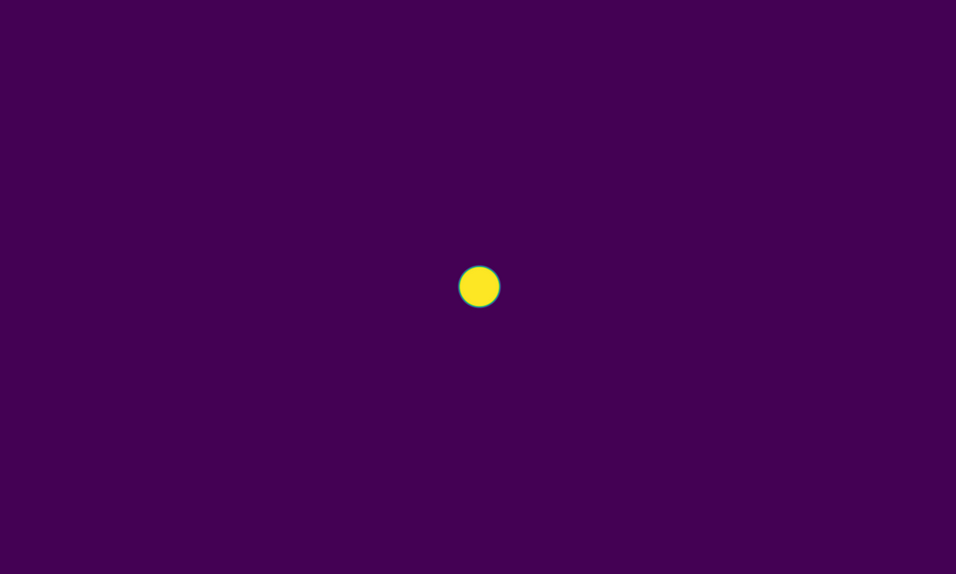}
    \caption{An example of a disk-shaped PSF. As it is a rasterized approximation, there is a small linear slope around the edges to avoid nonsensical discontinuities.}
    \label{fig:psf-disk}
\end{figure}

The PSF may also be defined using Zernike polynomial aberration representation, if the polynomial parameters are empirically determined. These are polynomials defined on the unit disk that are usually used to characterize refractive aberrations in wavefront aberrometers, and can be used to account for aberrations other than simple defocus. The aberration function is defined as the sum of the Zernike polynomial individual aberrations that compose it, as \(\mathfrak{Z}(x_\alpha,y_\alpha)=\sum_j Z^{j}(x_\alpha,y_\alpha)\), where \(Z^j\in\mathcal{Z}\) are Zernike polynomials; \((x_\alpha,y_\alpha)\) are coordinates on the pupil aperture plane. Fourier optics tells us that the PSF must thus be given by the Fourier transform:
\begin{align}
    k(x,y)\coloneqq\left|\iint_{-\infty}^{+\infty} {A(x_\alpha,y_\alpha)\cdot\mathsf{e}^{-i2\lambda^{-1}\pi\mathfrak{Z}}}\cdot\mathsf{e}^{-i(xx_\alpha+yy_\alpha)}\;\mathrm{d}x\,\mathrm{d}y\right|^2,
\end{align}
where \(A(x_\alpha,y_\alpha)\) is a pupil amplitude function that is unity within the pupil in the pupil aperture plan, and null everywhere else. 

One then defines the blurring of the original image \(I(x,y)\) to be:
\begin{align}
    \hat{B}(\omega{x},\omega{y}) \coloneqq \hat{I}(\omega{x},\omega{y})\cdot{}\hat{k}(\omega{x},\omega{y}) \\ 
    \implies B({x},{y}) = \mathcal{F}^{-1}\left\{\hat{I}(\omega{x},\omega{y})\cdot{}\hat{k}(\omega{x},\omega{y})\right\}.
\end{align}
Here, \(\omega\) is the angular frequency of the fourier tranforms, and \(\mathcal{F}\) is the fourier transform operation.

From here, we may define a function \(h(x,y)\) that inverts the operation above, such that:
\[
\mathcal{F}^{-1}\left\{\hat{I}(\omega{x},\omega{y})\cdot{}\hat{k}(\omega{x},\omega{y})\cdot{}\hat{h}(\omega{x},\omega{y})\right\} = I(x,y).
\]
That is, \(h(x,y)\) and \(k(x,y)\) cancel each other out to give the \emph{perfect PSF}, that is a bright single pixel in the origin. We define \(h(x,y)*k(x,y)=\mathbbm{1}(x,y)\), where \(\mathbbm{1}\) denotes the identity. Note that:
\[
k(x,y)*h(x,y)\equiv \mathcal{F}^{-1}\left\{\hat{k}(\omega{x},\omega{y})\cdot{}\hat{h}(\omega{x},\omega{y})\right\},
\]
and \(*\) is the convolution operator.

The recovery of \(I(x,y)\) from \(B(x,y)\) is the process of \emph{deconvolution}. This is not as simple as it seems, for in defining \(h(x,y)\) as follows:
\[
\hat{h}(\omega x,\omega y)=\begin{cases}
\hat{k}(\omega x,\omega y)^{-1} & \text{if $x\geq \epsilon$}, \\
0 & \text{otherwise},
\end{cases}
\]

We introduce severe artefacts in the convolution (that is a deconvolution by \(k(x,y)\)), and in the introduction of negative pixel intensities.

Here we note that by deconvolving an on-screen image, \(I(x,y)\), by the PSF of the human eye, \(k(x,y)\) will, theoretically, define a pre-distorted image \(P(x,y)\), such that the defocused eye perceives it clearly through the convolution \(P(x,y)*k(x,y)=I(x,y)\), that is the original image.

Recall from (1) that \(r\) was the variable upon which the PSF is dependent. We take \(r\) to be:
\[
r=a\frac{\left|d_{f}-d_{0}\right|}{d_{f}}.
\]
Here, $a$ is the pupil diameter; $f$ is the focal length; $d_{o}$ is the distance from the display; $d_{f}$ is the distance to the plane of focus. $d_{f}$ can be computed using the thin lens equation:
$$
\frac{1}{d_{f}}+\frac{1}{d_{e}}=\frac{1}{f},
$$
where $d_{e}$ is the eye’s internal diameter.

In the created prototype, we have made use of Wiener deconvolution to minimize noise.

As the user is susceptible to movement while viewing the screen, they will continue to view the screen with PSF \(k(x,y)\), but the on-screen PSF must be perspective transformed to ensure that spatial and angular ratios are conserved from all angles---in the case of the circle of confusion, it must appear circular from all angles, for the deconvolution to be effective. We use the coordinates, and dimensions of the eye to create a 2D transformation matrix encoding the perspective transformation to the PSF's current perception. We then apply the left inverse matrix to the original PSF, obtaining the perspective-corrected image.

This is then integrated with live distance and angle tracking, so that at discrete intervals, the PSF will be updated to reflect changes in distance and angle from the screen.

\subsection{User Interface}
\begin{figure}
    \centering
    \includegraphics[width=\textwidth]{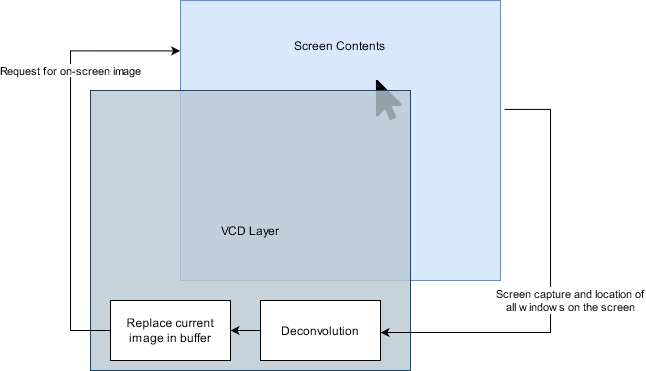}
    \caption{The front VCD layer and the back screen contents layer exchanging data. The VCD layer itself is a double buffered surface.}
    \label{fig:vcd-layer}
\end{figure}

We take in the user's optical prescription in the beginning, assuming simple defocus, so that we can approximate the PSF using (1). We then create a layer that continuously screen captures, and feeds it as an input to the deconvolution algorithm at discrete intervals. The output of the deconvolution algorithm is then used as a buffer for what is displayed on the VCD layer, as seen in Figure~\ref{fig:vcd-layer}. We use double buffering to reduce lag, wherein we treat the VCD layer as a buffer (the \emph{front buffer}) whose contents are instantly updated after deconvolution is complete.

The image whose deconvolution is in progress is stored in a separate buffer (the \emph{back buffer}), and the two buffers are swapped after processing is complete; hence the name \emph{double} buffering~\cite{doublebuffering}.
The alternative is to hide and re-display the layer as screen captures are taken and deconvolution is computed, which results in flickering.

\subsection{Live User Location Tracking}
\begin{figure}[H]
    \centering
    \includegraphics[width=0.5\linewidth]{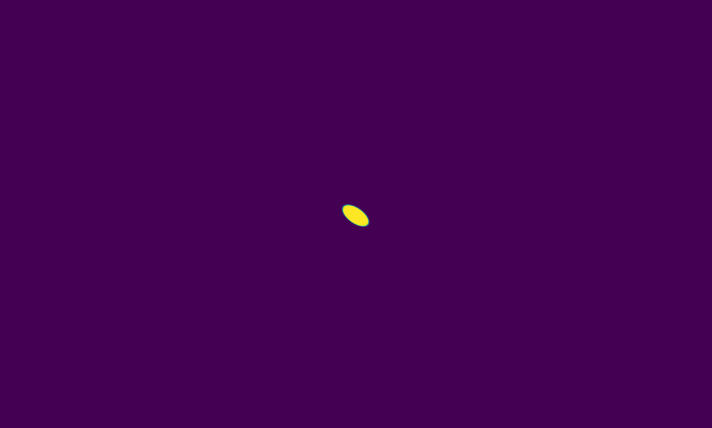}
    \caption{The perspective transformed Point Spread Function at a 45$^{\circ}$ angle.}
    \label{fig:perspectivetransformpsf}
\end{figure}

\ptitle{PSF Approximation With Respect To Angle}
To ensure real-time efficacy for users of the vision-correcting display, it is imperative that the device functions optimally across diverse viewing angles. Achieving the predistortion of the PSF can be effectively realized through a perspective transformation, by the following steps:
\begin{enumerate}
    \item Normalize the input image (that is, the PSF in Figure~\ref{fig:psf-disk}) to a grayscale format and scale pixel values.
    \item Represent 2D image points as homogeneous 3D coordinates.
    \item Define the projection matrix. The third column of the projection matrix translates the coordinates, shifting the origin to the center of the image 
    \[
P = \begin{bmatrix}
  1 & 0 & -\frac{c}{2} \\
  0 & 1 & -\frac{r}{2} \\
  0 & 0 & 0 \\
  0 & 0 & 1 \\
\end{bmatrix}
\]
\item Define the rotation matrix with respect to \(x\)- and \(y\)-axes ($R_{x}$ and $R_{y}$). The former simulates the effect of tilting or rotating the camera or viewing perspective along the horizontal axis. The rotation is defined by angle $(\theta_x)$. The off-diagonal elements $-\sin(\theta_x)$ and $\sin(\theta_x)$ introduce the rotation effect. The cosine terms $\cos(\theta_x)$ preserve spatial relationships during the rotation. As $\theta_x$ increases, the cosine term determines how much of the original \(y\)-coordinate is preserved during the rotation. This is similar in the $R_{y}$ vertical tilt matrix as well. 
\[
R_{x} = \begin{bmatrix}
  1 & 0 & 0 & 0 \\
  0 & \cos(\theta_x) & -\sin(\theta_x) & 0 \\
  0 & \sin(\theta_x) & \cos(\theta_x) & 0 \\
  0 & 0 & 0 & 1 \\
\end{bmatrix}
\]
\[
R_{y} = \begin{bmatrix}
  \cos(\theta_y) & 0 & \sin(\theta_y) & 0 \\
  0 & 1 & 0 & 0 \\
  -\sin(\theta_y) & 0 & \cos(\theta_y) & 0 \\
  0 & 0 & 0 & 1 \\
\end{bmatrix}
\]

\item Define a translation matrix $T$ for translation with respect to the \(z\)-axis---this corresponds to moving the camera or the viewpoint forward or backward. The translation is applied to the homogeneous coordinate, introducing a shift along the depth (\(z\)) axis. The third column ensures no change in the $z$-coordinate, except for the translation along the \(z\)-axis represented by $\xi$. The matrix is as follows:
\[
T = \begin{bmatrix}
  1 & 0 & 0 & 0 \\
  0 & 1 & 0 & 0 \\
  0 & 0 & 1 & \xi \\
  0 & 0 & 0 & 1 \\
\end{bmatrix}
\]

\item Define the camera intrinsics matrix $A_2$. Here, $f$ refers to focal length, determined based on the desired field of view. The principal point on the \(x\)- and \(y\)-axes is defined by the midpoint. The third row is fixed for homogeneous coordinates. $A_2$ converts 3D coordinates to 2D image coordinates, incorporating camera intrinsics. The FOV used in the prototype is \(80^\circ\)---the larger the FOV, the wider the perspective captured. \(c,r\) are the columns and rows.
\[
A_2 = \begin{bmatrix}
  f & 0 & \frac{\text{c}}{2} & 0 \\
  0 & f & \frac{\text{r}}{2} & 0 \\
  0 & 0 & 1 & 0 \\
\end{bmatrix}
\]

\item The matrices are multiplied in this order to achieve the desired perspective transformation. Transformation of the image by the perspective transformation matrix results in the pre-distorted output PSF, adapting the image to different viewing angles. Finally, the warped image is normalized, ensuring pixel values are within a valid intensity range.
\[
(A_2 T  R_y  R_x A_1)K = K_t,
\]
where \(K\) is the discretized PSF \(k(x,y)\) in matrix form, resulting in Figure~\ref{fig:perspectivetransformpsf}.

\end{enumerate}

\begin{figure}[H]
    \centering
    \includegraphics[width=0.5\linewidth]{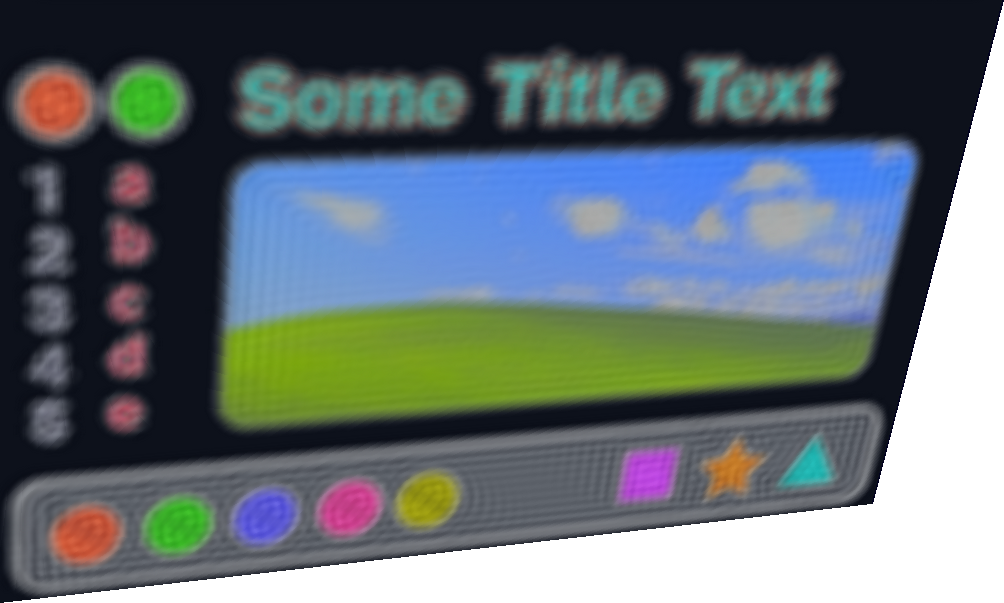}
    \caption{The display with the deconvolved image relative to the user (45$^{\circ}$ angle).}
    \label{fig:angleperception}
\end{figure}

\begin{figure}[H]
    \centering
    \includegraphics[width=0.5\linewidth]{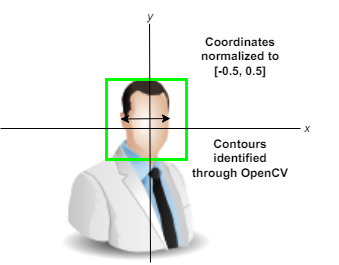}
    \caption{Representation of face recognition implementation}
    \label{fig:contours}
\end{figure}

\ptitle{Live Distance Estimation through Webcam}
The algorithm for live calculation of the user's distance from the screen involves an infinite loop, wherein each frame is captured, converted to RGB, and face locations are identified using \texttt{dlib}'s face recognition model. This creates contours as seen in Figure~\ref{fig:contours}. The model used has an accuracy of 99.38\% on the Labeled Faces in the Wild benchmark. 

The width of the face in pixels is set as a constant value. Then, using the horizontal FOV of the camera, the distance is calculated based on the principle that the user's face and its image on the live feed form similar triangles. 
\begin{align*}r = \frac{u_{w} f_{w}}{2 f_{p} \tan\left(\frac{\textsc{fov}_{h}}{2}\right)},\end{align*}
where $r$ is the absolute distance, $u_{w}$ is the user's face width in meters, $f_{w}$ is the frame width, $f_{p}$ is the face width in pixels as it appears in the live feed, and $FOV_{h}$ is the horizontal FOV of the camera. 

The calculated distance \(r\) is the estimated distance of the user from the screen in the given frame.
It is to be acknowledged that the distance calculation may deviate from the actual distance. If advanced depth sensing technologies such as Time-of-Flight (ToF) cameras are used, where the time taken for emitted light to travel to the user and back is measured, precise  information can be gathered. This would surpass traditional methods by mitigating the impact of ambient lighting conditions and ensuring robust performance across diverse environments. An alternative approach involves utilizing multiple sensors in a stereo vision setup enhances accuracy further by triangulating the user's position. It is worth noting that such advanced setups require sophisticated hardware and computational resources, and does not meet the criteria for a live display that the layman should be able to use.

\ptitle{Live Horizontal and Vertical Angle Measurement Through Webcam}
The algorithm for live calculation of the user's horizontal and vertical angles occurs simultaneously. Within the algorithm, the user's face is localized in each frame, and the horizontal and vertical angles are computed based on the face's midpoint position within the frame. The horizontal angle is calculated by normalizing the face's midpoint deviation from the center of the frame, while the vertical angle is similarly determined using the midpoint's deviation from the frame's vertical center. These angles are then converted from degrees to radians, and the resulting data is stored.

\subsection{Real-Time Video Deconvolution}
The vision correcting display works on the principle of deconvolving singular images and displaying this to the user. Given that a video is merely a collection of frames, we have adapted our algorithm to work for video inputs as well. The prototype player allows interactive navigation through a video, including play/pause functionality, seeking, and caching frames for smoother playback. The player utilizes multi-threading for parallelization, with one thread handling caching, and another for rendering the frames and managing user interactions. When a video is inputted, upcoming frames (approximately 3 seconds' worth) are deconvolved and cached. This allows for efficient playback of videos. On an i9 32 GB RAM computer, it took 20 seconds to deconvolve 30 seconds of video or 1800 frames (90 FPS), whilst 27 FPS was observed on an i7 16GB RAM computer.

\subsection{Performance Optimization by Tiling}
The Wiener deconvolution algorithm runs with \(\mathcal{O}(n\log n)\) complexity, where \(n\) are the data size. The deconvolution algorithm's time complexity is dominated by the Fast Fourier Transform (FFT) and Inverse FFT (IFFT) operations used in its computation.

By slicing the original image into tiles, the 2D data size \(n\) is quadratically reduced, resulting in a much lower time complexity per individual tile (e.g., for division into tiles as powers of 2, \(\mathcal{O}\left(\frac{n}{2^a}\log\frac{n}{2^a}\right)\)). For the total number of tiles, the order of complexity becomes \(\mathcal{O}\left({n}\log\frac{n}{2^a}\right)\), which is much lower than the original time complexity. The tiles are then easily stitched back together---for instance, using basic array manipulation.

\subsection{Ringing Artefact Reduction}

We notice multiple ringing artefacts in a traditionally deconvolved image. A technique to remove majority of them that are not in the focus of the image was previously established in \cite{doi:10.1080/17415970802082823}, as follows.

First, the blurred on-screen image \(I(x,y)*k(x,y)\) is edge detected. An edge detection mask with a black fill, \(\mathbf{M}(x,y)\) is produced. The mask is then applied to the deconvolved image \(I(x,y)*h(x,y)\), producing \(\mathbf{M}(x,y)\otimes (I(x,y)*h(x,y))\). \(\otimes\) denotes logical intersection, and \(\oplus\) denotes logical conjunction. The inverse mask \(\mathbf{M}^{-1}(x,y)\) is applied to the original image as \(\mathbf{M}^{-1}(x,y)\otimes I(x,y)\), and the two images are superimposed as:
\[
\left(\mathbf{M}^{-1}(x,y)\otimes I(x,y)\right)\oplus\left[\mathbf{M}(x,y)\otimes (I(x,y)*h(x,y))\right].
\]
This is the image then displayed on-screen.

We introduce an intermediate step, where each edge-detected segment is analysed for text presence. If text is present, we deconvolve that segment separately, with a higher regularization constant, as it has more edges. We obtain, thus:
\[
\left[\mathbf{M}^{-1}(x,y)\otimes I(x,y)\right]\oplus\left(\bigoplus_{\partial{I}}\begin{cases}
    \partial I * h(x,y,\rho_1) & \text{if \(t(\partial I)=\varnothing\)}, \\
    \partial I * h(x,y,\rho_2) & \text{otherwise}.
\end{cases}\right),
\]
as the on-screen image, where \(\rho_2>\rho_1\), and \(t(I)\) returns the text present in a given image (\(\emptyset\equiv\texttt{""}\)). In this case, the function is supplied by Tesseract OCR models. \(\partial I\) are the individual edge detected and masked components of the image, such that:
\[
\bigoplus_{i=1}^n\partial_i{I}=I(x,y)
\]

\subsection{Choice \& Optimization of Colour Space}

The question arises of how a colored image may be pre-distorted, as only monochannel images have been considered till now. For colour encoding, RGB is typically used. The original image may be represented in the form:
\[
I(x,y) = (r(x,y), g(x,y), b(x,y)),
\]
with a colour tuple \((r,g,b)\) associated with each coordinate. We may thus apply convolution or deconvolution operations to the individual color channel functions \(r,g,b:\mathbb{Z}^3\to \mathbbm{clr}\), where \(\mathbbm{clr}\coloneqq \{\mu \mid 0\leq\mu\leq 255; \mu\in\mathbb{Z}\}\)

Deconvolution is a space-intensive process. The fast fourier transform alone is of space complexity \({\mathcal{O}(n\log n)}\), and the deconvolution operation's complexity is dominated by the fourier transform. It is thus infeasible to conduct deconvolution operations on each of the \(r,g,b\) channels, which triples the processing time and takes a high computing toll. Additionally, we observe severe color bleeding and color perversion in using \(r,g,b\) separate channel filtering. In the case of \(r,g,b\) channel deconvolution, the deconvolution is represented:
\begin{align}
    (r(x,y), g(x,y), b(x,y))\mapsto(r(x,y) * h(x,y), g(x,y) * h(x,y), b(x,y) * h(x,y)).
\end{align}

That is, three deconvolutions must take place; one for each of the channels. One may, instead, attempt to decompose the full-color image into a different set of channels, representative of different aspects of the color. For instance, YUV (in modern displays, this is YCbCr). Y is the luma channel---solely the luma channel will give a grayscale version of the image. U and V represent deficiency in red and blue respectively. Our decomposition becomes:
\[
I(x,y) = (y(x,y), u(x,y), v(x,y)),
\]
and we see that only the luma channel must be pre-distorted. Non-analytically, the U and V channels will retain their original colors, and the eye will percieve the colors to fit into the shapes defined in the luma channel. We note that \(y,u,v:\mathbb{Z}^3\to \mathbbm{clr}\); here, \(\mathbbm{clr}\coloneqq \{\mu \mid 0\leq\mu\leq 1; \mu\in\mathbb{R}\}\) as a matter of convention. Now, the deconvolution is given:
\begin{align}
    (y(x,y), u(x,y), v(x,y))\mapsto(y(x,y) * h(x,y), u(x,y), v(x,y)),
\end{align}
reducing the operation to a single deconvolution. We may alternatively use the LMS color space, which decomposes colors in a similar manner (one luma and two chroma channels), but while taking into account the human eye's perception to different color intensities. We notice a very subtle difference in the simulation, between the two color spaces.

In integrating this into the live vision correction, we use graphics pipelining to process multiple images simultaneously to avoid lag. Additionally, individual images were segmented into smaller tiles, that were then processed. As the complexity of deconvolution is of a greater order than linear, processing multiple smaller images takes less time than processing a single large image.

\subsection{Higher Order Aberrations}
\begin{figure}[H]
    \centering
    \includegraphics[width=0.5\textwidth]{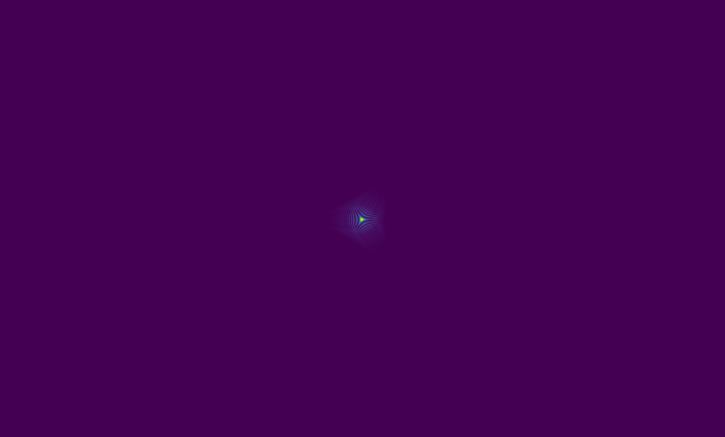}
    \caption{A sample oblique trefoil PSF.}
    \label{fig:oblique_trefoil}
\end{figure}

\begin{figure}[H]
    \centering
    \includegraphics[width=0.5\textwidth]{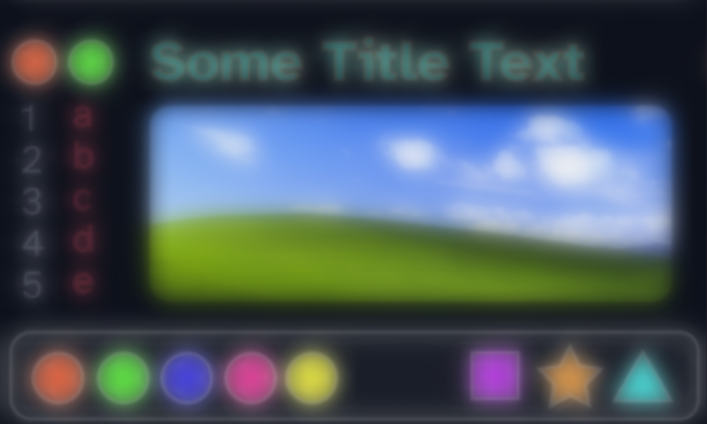}
    \caption{Non-trefoil affected user User’s perception of image without correction.}
    \label{fig:trefoil_withoutcorrection}
\end{figure}

\begin{figure}[H]
    \centering
    \includegraphics[width=0.5\textwidth]{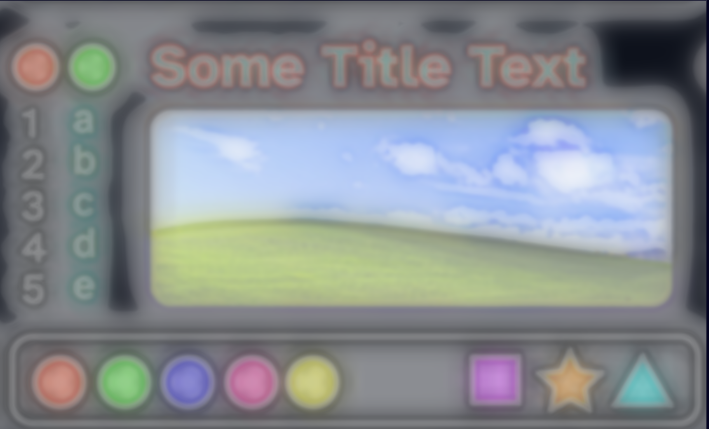}
    \caption{Perceived image for trefoil-afflicted user.}
    \label{fig:trefoil_perceived}
\end{figure}

Hartmann-Shack Wavefront Aberrometers give an empirical reading of the magnitude of different refractive aberrations—this can be used to reconstruct a PSF. As seen in the subsequent figures, the deconvolution algorithm also works for such non-circular PSFs. In Figure~\ref{fig:oblique_trefoil}, the PSF is non-circular (unlike Figure~\ref{fig:psf-disk}), resulting in a blurred perception of the original image with trefoil (Figure~\ref{fig:trefoil_withoutcorrection}. We show a clearer image for this higher order aberration (Figure~\ref{fig:trefoil_perceived}).

\section{Results}
Qualitatively, we see that the images retain bright colours in the reconvolved images. Likewise, the ringing artifacts degrading the image quality are masked, ensuring that the image contrast is not lost.

To assess the performance of our display, we compare the original image with the reconvolved image, which simulates how the corrected image appears to the viewer. We evaluate the quality of the reconvolved image using multiple metrics: Peak Signal-to-Noise Ratio (PSNR), Root Mean Square Error (RMSE), Absolute Error (AE), Normalized Cross-Correlation (NCC), and Structural Similarity Index (SSIM). These metrics provide insights into the effectiveness of our approach in mitigating visual aberrations. The following images will be utilized for comparison.

\begin{figure}[H]
    \centering
    \includegraphics[width=0.5\linewidth]{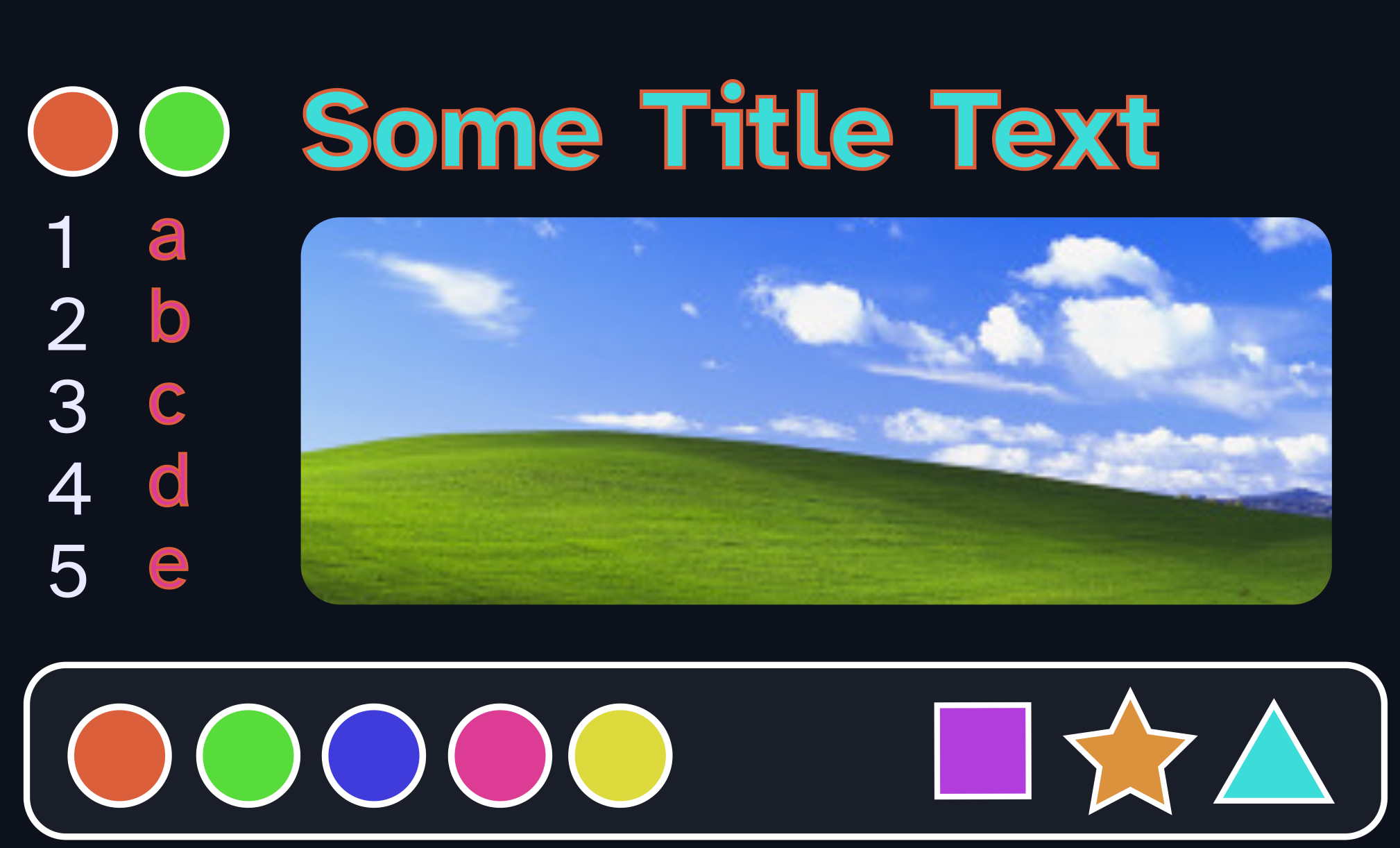}
    \caption{Original (reference) image}
    \label{fig:original_image}
\end{figure}
\begin{figure}[H]
    \centering
    \includegraphics[width=0.5\linewidth]{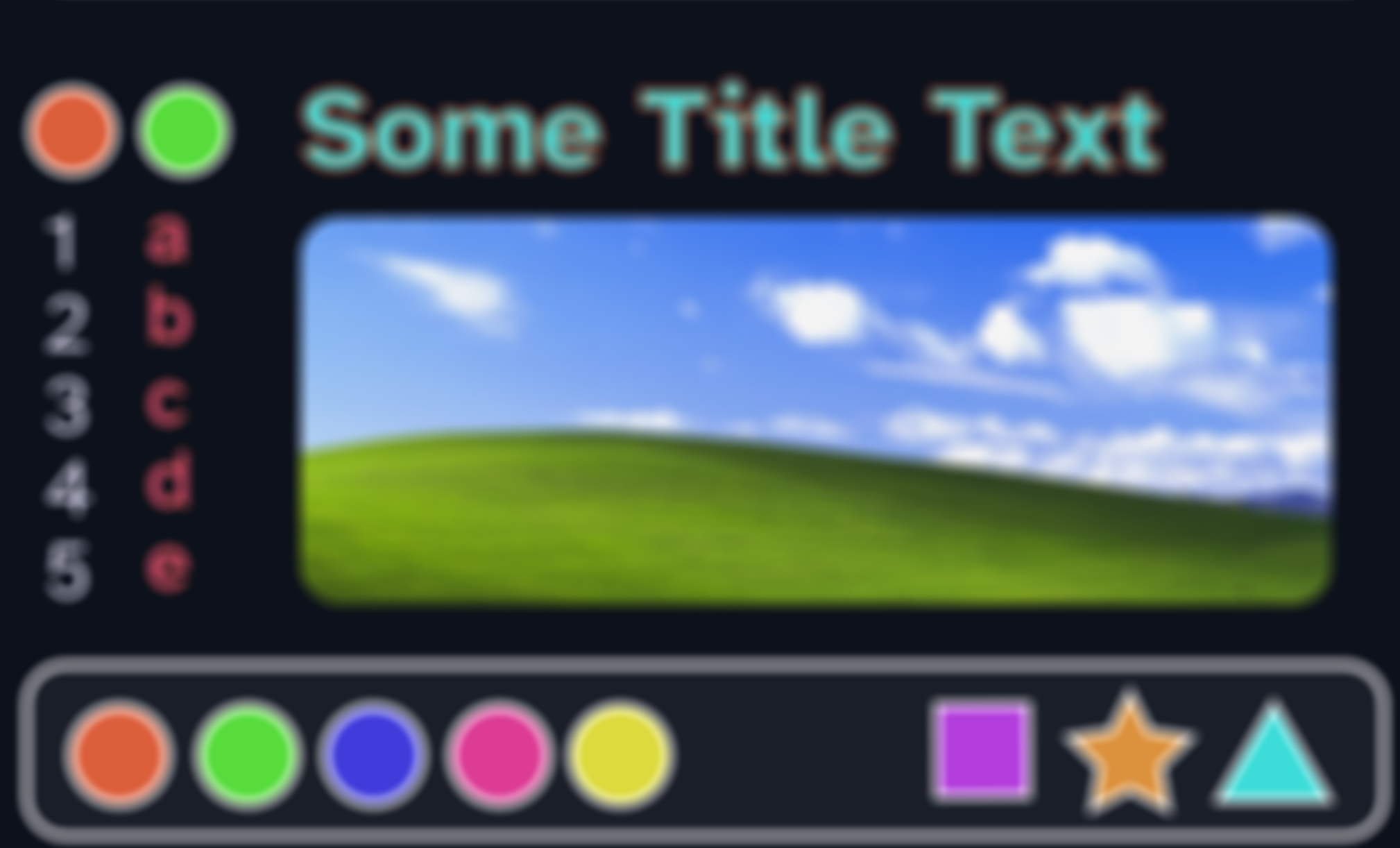}
    \caption{Blurred (perceived) image without correction}
    \label{fig:blurred_image}
\end{figure}
\begin{figure}[H]
    \centering
    \includegraphics[width=0.5\linewidth]{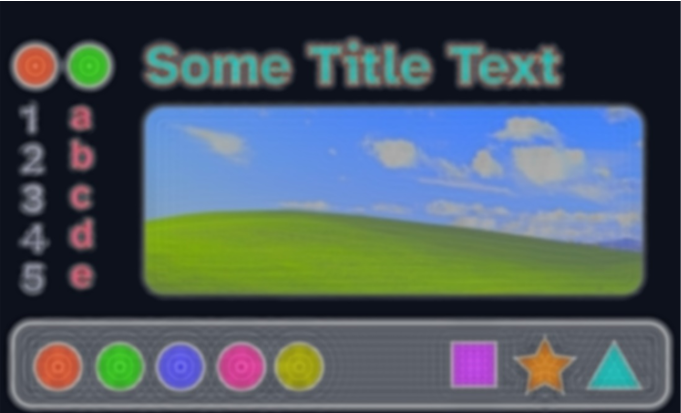}
    \caption{Reconvolved (perceived) image with correction}
    \label{fig:reconvolved_image}
\end{figure}

PSNR is a widely used metric for evaluating image quality by comparing the original and distorted images. It measures the ratio between the maximum possible signal value and the magnitude of noise introduced during image reconstruction. Higher PSNR values indicate better fidelity, implying that the reconvolved image closely resembles the original~\cite{hore2010image}. It is given:
\begin{align}
    \textsc{psnr} = 10 \;\log\left( \frac{\operatorname{max}(I(x,y))^2}{\overline{\mathrm{E_\text{squ}}}} \right)
\end{align}

where $I$ is the image, and $\overline{\mathrm{E_\text{squ}}}$ is the mean square error between the original and reconvolved images.

The PSNR was computed across different colour channels and collectively for the entire image, seen in Table \ref{tab:results}. These values show a moderate level of noise in the reconvolved image, suggesting areas for optimization in the deconvolution parameters to achieve a closer approximation to the original image quality. The consistency across channels also confirms that our approach maintains colour integrity, which is crucial for practical applications.

The Absolute Error (AE) measures the discrepancy between the corresponding pixel values of the original and reconvolved images. It is calculated as the sum of the absolute differences between the pixel intensities of the two images~\cite{hodson2022root}. AE for each pixel is defined as:\begin{align}
    E_\text{abs} = \texttt{m}\left(I(x,y)\right) \ominus \texttt{m}\left(\left[I(x,y)*h(x,y)\right]*k(x,y)\right),
\end{align}
where $I$ is the original image, \(\texttt{m}:\mathbb{Z}^3\to\mathbb{Z}\) is the monochromatic filter function. As usual, \(k,\,h\) respectively denote the PSF and IPSF. Here, \(\cdot\ominus\cdot\) is naturally a component-wise subtraction operation.

\begin{figure}[H]
    \centering
    \includegraphics[width=0.5\textwidth]{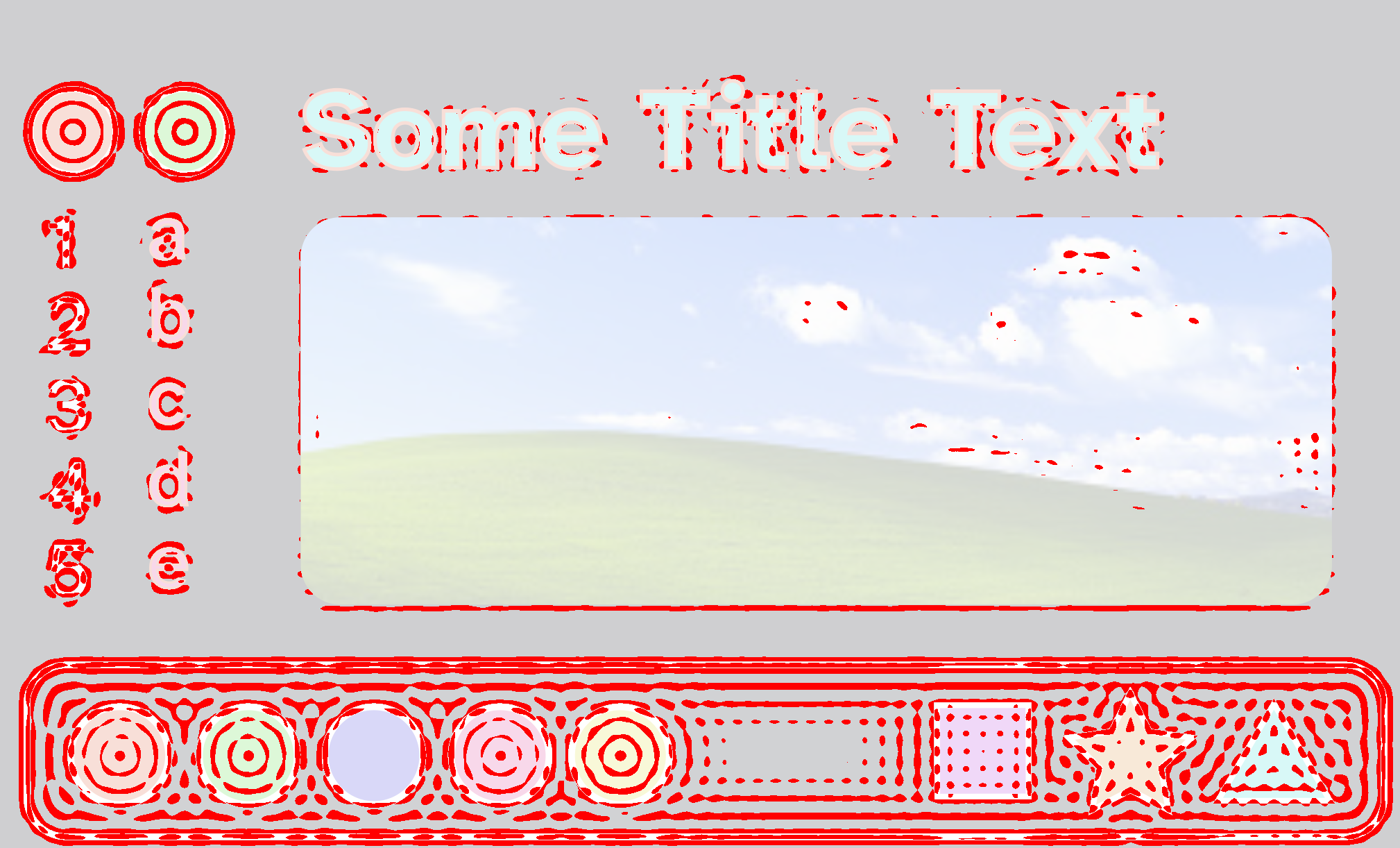}
    \caption{Highlighting pixel differences. It can be seen that differences primarily exist at the edges of the image.}
    \label{fig:pixelwisedifference}
\end{figure}

AE was computed separately for each colour channel as well as collectively for the entire image. This showed a level of variance in pixel values across all channels, with a percentage difference of 10.67\% from the original image. However, this is not a demonstration of perceived clarity, and is simply a pixel-wise difference. In Figure~\ref{fig:pixelwisedifference}, is observed that most differences primarily exist in edge regions.

The Structural Similarity Index (SSIM) is a metric used to assess the perceived quality of digital images and videos. SSIM is considered superior to simpler metrics such as Mean Squared Error (MSE) and PSNR because it incorporates perceptual phenomena, including texture, luminance, and contrast, which are critical in human visual perception~\cite{sara2019image}.
SSIM considers the closeness of the image luminance, which reflects the average brightness, it evaluates the similarity in contrast, which involves measuring the standard deviation of the pixel intensities, and it assesses the correlation coefficient between the two images, which quantifies how structural elements in the images correlate~\cite{ndajah2010ssim}. It is given:
\begin{align}
    \textsc{ssim}(x, y) = \frac{(2 \mu_x \mu_y + c_1)(2 \sigma_{xy} + c_2)}{(\mu_x^2 + \mu_y^2 + c_1)(\sigma_x^2 + \sigma_y^2 + c_2)},
\end{align}
where \( \mu_x \) and \( \mu_y \) represent the mean intensity values of images \( x \) and \( y \), \( \sigma_x \) and \( \sigma_y \) are the standard deviations, and \( \sigma_{xy} \) is the covariance of the two images. The constants \( c_1 \) and \( c_2 \) are included to stabilize the division with a weak denominator.

In Figure~\ref{fig:original_image} and Figure~\ref{fig:reconvolved_image}, the SSIM value achieved was 83.04\%, indicating a high degree of similarity between the original and the corrected images viewed through the display. 

\begin{longtable}[c]{|l|l|l|l|l|}
\hline
                             & Our Work & Huang, 2014 ~\cite{Huang:2014} & Huang, 2012 ~\cite{huang2012correcting}& Pamplona, 2012 ~\cite{pamplona2012tailored} \\ \hline
\endfirsthead
\endhead
SSIM                         & 83.04\%  & 50.59\%            & 27.89\%            & 47.28\%               \\ \hline
PSNR (RGB)                   & 14.3268  & 15.1538            & 8.4279             & 15.1366               \\ \hline
PSNR (\% diff.) & 0.00\%   & 0.03\%             & 0.02\%             & 0.03\%                \\ \hline
NCC (Coefficient)            & 0.742    & 0.423              & 0.206              & 0.380                 \\ \hline
Contrast                     & 100\%    & 45\%               & 15\%               & 100\%                 \\ \hline
\caption{Comparison of our work with existing literature, using the aforementioned metrics. It can be seen that our deconvolution in YUV space preserves image contrast.}
\label{tab:results}\\
\end{longtable}

\section{Conclusion}
\subsection{Contributions}
Thus, we see that our work has enabled approximate vision correction for refractive aberrations (SSIM = 83.04\%). This can be made considerably more accurate for the end user by allowing for empirical PSF input from a wavefront aberrometer. We have successfully optimized the traditional deconvolution approach heavily for the real-time vision correction aim, making use of image tiling and parallel processing, whilst contributing to signal processing though YUV space deconvolution---this can be used in traditional deconvolution applications as well. These optimizations and novel contributions are a step forward in making vision correcting displays feasible solutions in the real world.

\subsection{Limitations}
The deconvolution approach does not work well for small images due to their rasterized nature---thus, the deconvolution should ideally be performed on a continuous vectorized image, as at small scales, ringing artefacts tend to dominate the image, obscuring relevant details.

We also note the presence of seams where the image has been cut during the tiling process---their presence is apparent only because of the ringing artefacts. To avoid this, the image may be broken up into tiles with some padding, so that the ringing artefacts around the edges do not reduce clarity.

Similarly, the tiling approach described does not work for higher-order aberrations, because the initial aberration is applied to the entire image, and may involve shifting elements from one tile to another, resulting in a misaligned reconvolved image. This may be avoided, and the processing time may be reduced signifiantly, if tiling is not used---rather, the image may be broken up based on the presence of contours around letters and images. These can later be individually overlayed onto an undistorted background instead of separately tiling and masking.

\subsection{Improvements}
Improvements to this implementation could include the creation of a parallax barrier to separate left and right eye perception to accommodate different left and right power. In addition, very limited tests have been conducted with human subjects. We aim to begin testing the display on patients with clear vision, after which their ciliary muscle can be relaxed and myopia can be induced via lenses. This will accurately match the point spread function entered into the display.

\begin{acknowledgments}
We thank Professor Partha Pratim Das from the Department of Computer Science at Ashoka University for his invaluable counsel and guidance in this project. His expertise in image processing helped us refine our implementation and increase its applicability. It was with his support that we presented our work at the International Science and Engineering Fair, 2024. We are immensely grateful to Ashoka University and Mphasis Limited for giving us the opportunity to carry forward our work and for offering us the mentorship without which this project would not have been possible.
\end{acknowledgments}
\clearpage
\nocite{*}
\bibliographystyle{ieeetr}
\bibliography{bibliography}

\end{document}